\documentclass[]{emulateapj}
\usepackage{graphicx}
\usepackage{graphics}
\usepackage{float}
\usepackage{hyperref}
\usepackage{amsmath}

\usepackage{epstopdf}

\usepackage{natbib}

\usepackage{txfonts}
\bibpunct{(}{)}{;}{a}{,}{,}

\makeatletter
\long\def\@makecaption#1#2{%
  \vskip\abovecaptionskip
  \sbox\@tempboxa{#1: #2}%
  \ifdim \wd\@tempboxa >\hsize
    #1: #2\par
  \else
    \global \@minipagefalse
    \hb@xt@\hsize{\box\@tempboxa\hfil}%
  \fi
  \vskip\belowcaptionskip}
\makeatother

\shorttitle{The stellar IMF at very low metallicities}
\shortauthors{Dopcke et al.}

\begin{document}

\title{The stellar IMF at very low metallicities}

\author{Gustavo Dopcke, Simon C. O. Glover, Paul C. Clark and Ralf S. Klessen}
\affil{Zentrum f\"{u}r Astronomie der Universit\"{a}t Heidelberg, Institut f\"{u}r Theoretische Astrophysik, Albert-Ueberle-Str. 2, 69120 Heidelberg, Germany}
\email{gustavo@uni-hd.de}

\begin{abstract}
The theory for the formation of the first population of stars (Pop III) predicts an initial mass function (IMF) dominated by high-mass stars, in contrast to the present-day IMF, which tends to yield mostly stars with masses less than 1 $M_{\odot}$. The leading theory for the transition in the characteristic stellar mass predicts that the cause is the extra cooling provided by increasing metallicity. In particular, dust can overtake H$_2$ as the leading coolant at very high densities.
The aim of this work is to determine the influence of dust cooling on the fragmentation of very low metallicity gas. To investigate this, we make use of high-resolution hydrodynamic simulations with sink particles to replace contracting protostars, and analyze the collapse and further fragmentation of star-forming clouds. We follow the thermodynamic response of the gas by solving the full thermal energy equation, and also track the behavior of the dust temperature and the chemical evolution of the gas. We model four clouds with different metallicities (10$^{-4}$, 10$^{-5}$, 10$^{-6} \rm$ Z$_{\odot}$, and 0 ), and determine the properties of each cloud at the point at which it undergoes gravitational fragmentation.
We find evidence for fragmentation in all four cases, and hence conclude that there is no critical metallicity below which fragmentation is impossible. Nevertheless, there is a clear change in the behavior of the clouds at Z $= 10^{-5}$ Z$_{\odot}$, caused by the fact that at this metallicity, fragmentation takes longer to occur than accretion, leading to a flat mass function at lower metallicities.
\end{abstract}


\section{Introduction}\label{int}
The first burst of star formation in the Universe was thought to give rise to massive stars, the so-called Population\,III (Pop.~III), with numerical simulations predicting masses in the range 20-150 $M_{\odot}$ \citep[e.g.][]{2002Sci...295...93A, 2002ApJ...564...23B, 2007ApJ...654...66O, 2008Sci...321..669Y}. However, recent results show that lower mass stars can also be formed, albeit with characteristic masses above the solar value \citep{2011Sci...331.1040C, 2011ApJ...727..110C,2011ApJ...737...75G, 2010MNRAS.403...45S,2011arXiv1109.3147S, 2011MNRAS.414.3633S, 2012arXiv1202.5552G}.
This contrasts with present-day star formation, which typically yields stars with masses less than 1 $M_{\odot}$ \citep{2002Sci...295...82K, 2003PASP..115..763C}, and so at some point in the evolution of the Universe there must have been a transition from primordial (Pop.\,III) star formation to the mode of star formation we see today (Pop.\,II/I).

When gas collapses to form stars, gravitational energy is transformed into thermal energy and unless this can be dissipated in some fashion, the collapse will come to a halt. Thermal energy can be removed by processes such as atomic fine structure line emission, molecular rotational or vibrational line emission, or thermal emission from dust grains. In some cases, these processes are able to cool the gas significantly during the collapse. This temperature drop can promote gravitational fragmentation \citep{2004RvMP...76..125M, 2007prpl.conf..149B} by decreasing the Jeans mass, which means that instead of forming very massive clumps, with fragment masses corresponding to the initial Jeans mass in the cloud, it can instead form a large number of fragments with lower masses.
  
If the gas is cooled only by molecular hydrogen emission, numerical simulations show that most of the stellar mass would be in massive stars \citep{1999ApJ...515..239N,2001ApJ...548...19N,2002ApJ...569..549N,2002Sci...295...93A, 2002ApJ...564...23B, 2007ApJ...654...66O, 2008Sci...321..669Y, 2011MNRAS.414.3633S}.
This happens because H$_2$ cooling becomes inefficient for temperatures below 200K and densities above $10^4 \rm cm^{-3}$. At this temperature and density, the mean Jeans mass at cloud fragmentation is 1,000 times larger than in present-day molecular clouds.

Metal line cooling and dust cooling are effective at lower temperatures and larger densities, and so it has been proposed that metal enrichment of the interstellar medium by previous generations of stars causes the transition from Pop.\,III to Pop.\,II. This suggests that there might be a critical metallicity Z$_{\rm crit}$ at which the mode of star formation changes.

The main coolants that have been studied in the literature are C \textsc{ii} and O \textsc{i} fine structure emission \citep{2001MNRAS.328..969B, 2003Natur.425..812B,  2006ApJ...643...26S, 2007MNRAS.380L..40F, 2009ApJ...696.1065J, 2009ApJ...694.1161J, 2007ApJ...661L...5S, 2009ApJ...691..441S}, and dust emission \citep[e.g.][]{2002ApJ...571...30S,2006MNRAS.369.1437S,2012MNRAS.419.1566S, 2005ApJ...626..627O,2010ApJ...722.1793O}. Carbon and oxygen are identified as the key species because in the temperature and density conditions that characterize the early phases of Pop.\,III star formation, the O \textsc{i} and C \textsc{ii} fine-structure lines dominate over all other metal line transitions \citep{1989ApJ...342..306H}. By equating the C \textsc{ii} or O \textsc{i} fine structure cooling rate to the compressional heating rate due to free-fall collapse, one can define critical abundances $[\rm C/H] = -3.5$ and $[ \rm O/H] = -3.0$\footnote{$[\rm X/ \rm Y] = log_{10}(N_{\rm X}/N_{\rm Y})_{\star} - log_{10}(N_{\rm X}/N_{\rm Y})_{\odot}$, for elements X and Y, where $\star$ denotes the gas in question, and where $\rm N_X$ and $\rm N_Y$ are the mass fractions of the elements X and Y.} for efficient metal line cooling \citep{2003Natur.425..812B}.

If one assumes that effective fine-structure cooling is a necessary condition for the formation of Population\,II stars, then all such stars should have a ``transition discriminant" $D_{\rm trans} \equiv \log(10^{[\rm C/H]}+0.3 \times 10^{[\rm O/H]})$ greater than a critical value $D_{\rm trans, crit} = -3.5$ \citep{2007MNRAS.380L..40F}. Although most metal-poor stars lie above this value, at least one star has been observed to lie below it \citep[SDSS J102915+172927;][]{2011Natur.477...67C}, and there are other objects that might also have values of $D_{\rm trans}$, below the critical value: CS30336-049 \citep{2008ApJ...681.1524L} and Scl07-50 \citep{2010A&A...524A..58T}.

Previous works \citep{2009ApJ...696.1065J, 2009ApJ...694.1161J} have shown that this metallicity threshold does not represent a critical metallicity. The fact that the metal-line cooling rate has a larger value than the compressional heating rate does not necessarily lead to fragmentation, and even in cases where it does, the fragments that form have masses M $\gg 1$M$_{\odot}$. This points towards a different process leading to low-mass stars for the early Universe \citep{2012MNRAS.tmp.2494K}.

A more promising way to form low mass Pop.\,II stars involves dust cooling. Dust cooling models \citep[e.g.][]{2005ApJ...626..627O, 2010ApJ...722.1793O, 2011ApJ...729L...3D, 2006MNRAS.369.1437S, 2012MNRAS.419.1566S} predict a much lower critical metallicity (Z$_{\rm crit} \approx 10^{-4} - 10^{-6}$ Z$_{\odot}$), with most of the uncertainty coming from the nature of the dust in high-redshift galaxies.

At densities $n \gtrsim 10^{11}$ cm$^{-3}$ dust cooling becomes efficient \citep{2010ApJ...722.1793O}, since inelastic gas-grain collisions are more frequent \citep{1979ApJS...41..555H}. This cooling enhances fragmentation, and since it occurs at high densities, the distances between fragments can be very small \citep{2000ApJ...534..809O, 2005ApJ...626..627O, 2002ApJ...571...30S, 2006MNRAS.369.1437S, 2010MNRAS.402..429S}. In this regime, interactions between fragments will be common, and analytic models of fragmentation are unable to predict the mass distribution of the fragments. A full 3D numerical treatment, following the fragments, is needed.

Initial attempts at modeling fragmentation in low metallicity gas were made by \cite{2006ApJ...642L..61T,2008ApJ...676L..45T} and \cite{2008ApJ...672..757C}. These studies described the thermal evolution of the gas using effective equations of state derived from the one-zone calculations of \cite{2005ApJ...626..627O}, and showed that the cooling provided by dust does indeed lead to fragmentation. This treatment assumes, however, that the gas temperature adjusts instantaneously to a new equilibrium whenever the density changes and hence ignores thermal inertia effects. This may yield too much fragmentation.

In \cite{2011ApJ...729L...3D}, we improved upon these previous treatments by solving the full thermal energy equation, and calculating the dust temperature through the energy equilibrium equation. We assumed that the only significant external heat source is the cosmic microwave background (CMB), and included its effects in the calculation of the dust temperature. We found that model clouds with metallicities as low as 10$^{-4}$ Z$_{\odot}$ or 10$^{-5}$ Z$_{\odot}$ do indeed show evidence for dust cooling and fragmentation, supporting the predictions of \cite{2006ApJ...642L..61T,2008ApJ...676L..45T} and \cite {2008ApJ...672..757C}.

In this work, we simulate the evolution of star-forming clouds for a wider range of metallicities (10$^{-4}$, 10$^{-5}$, 10$^{-6}$ Z$_{\odot}$, and 0), and study the effect that this has on the mass function of the fragments that form. We also investigate how properties such as cooling and heating rates, and number of Bonnor-Ebert masses \citep{1956MNRAS.116..351B, 1955ZA.....37..217E} of the fragmenting clouds vary with metallicity and whether there is any systematic change in behavior with increasing metallicity.

\section{Simulations}

\subsection {Numerical method}\label{meth}
We model the collapse of a low-metallicity gas cloud using a modified version of the Gadget 2 \citep{2005MNRAS.364.1105S} smoothed particle hydrodynamics (SPH) code. To enable us to continue our simulation beyond the formation of the first very high density protostellar core, we use a sink particle approach \citep{1995MNRAS.277..362B, 2005A&A...435..611J}, in the same way as in \cite{2011ApJ...729L...3D}. Sink particles are created once the SPH particles are bound, collapsing, and within an accretion radius, $h_{acc}$, which we take to be 1.0 AU. The threshold number density for sink particle creation is $5.0 \times 10^{13} \rm cm^{-3}$. At the threshold density, the Jeans length at the minimum temperature reached by the gas is approximately one AU, while at higher densities the gas becomes optically thick and begins to heat up. Further fragmentation on scales smaller than the sink particle scale is therefore unlikely to occur. For further discussion of the details of our sink particle treatment, we refer the reader to \cite{2011ApJ...727..110C}.

We assume that the mean dust grain cross section is the same as for Milky Way dust and that the number density of dust grains is a factor Z/ Z$_{\odot}$ smaller than the Milky Way value \citep[see][]{2011ApJ...729L...3D}.
To treat the chemistry and thermal balance of the gas, we use the same approach as in \citet{2011ApJ...727..110C}, with the inclusion of dust cooling. The \citet{2011ApJ...727..110C} chemical network and cooling function were designed for treating primordial gas and do not include the chemistry of metals such as carbon or oxygen, or the effects of cooling from these atoms, or molecules containing them such as CO or H$_{2}$O. We justify this approximation by noting that previous studies of very low-metallicity gas \citep[e.g.][]{2005ApJ...626..627O, 2010ApJ...722.1793O} find that gas-phase metals have little influence on the thermal state of the gas. \cite{2010ApJ...722.1793O} showed that H$_2$O and OH are efficient coolants at $10^8 < n < 10^{10} \rm cm^{-3}$ for their one-zone model.
In their hydrodynamical calculations, however, the collapse is faster, and the effect of H$_2$O and OH is not perceptible. Therefore we do not expect oxygen-bearing molecules to have a noticeable effect on the thermal evolution of the gas.

For the metallicities and dust-to-gas ratios considered in this study, the dominant sources of cooling are the standard primordial coolants (H$_{2}$ bound-bound emission and collision-induced emission) and energy transfer from the gas to the dust.
\begin{figure}
  \centering
    \includegraphics[width=0.481\textwidth]{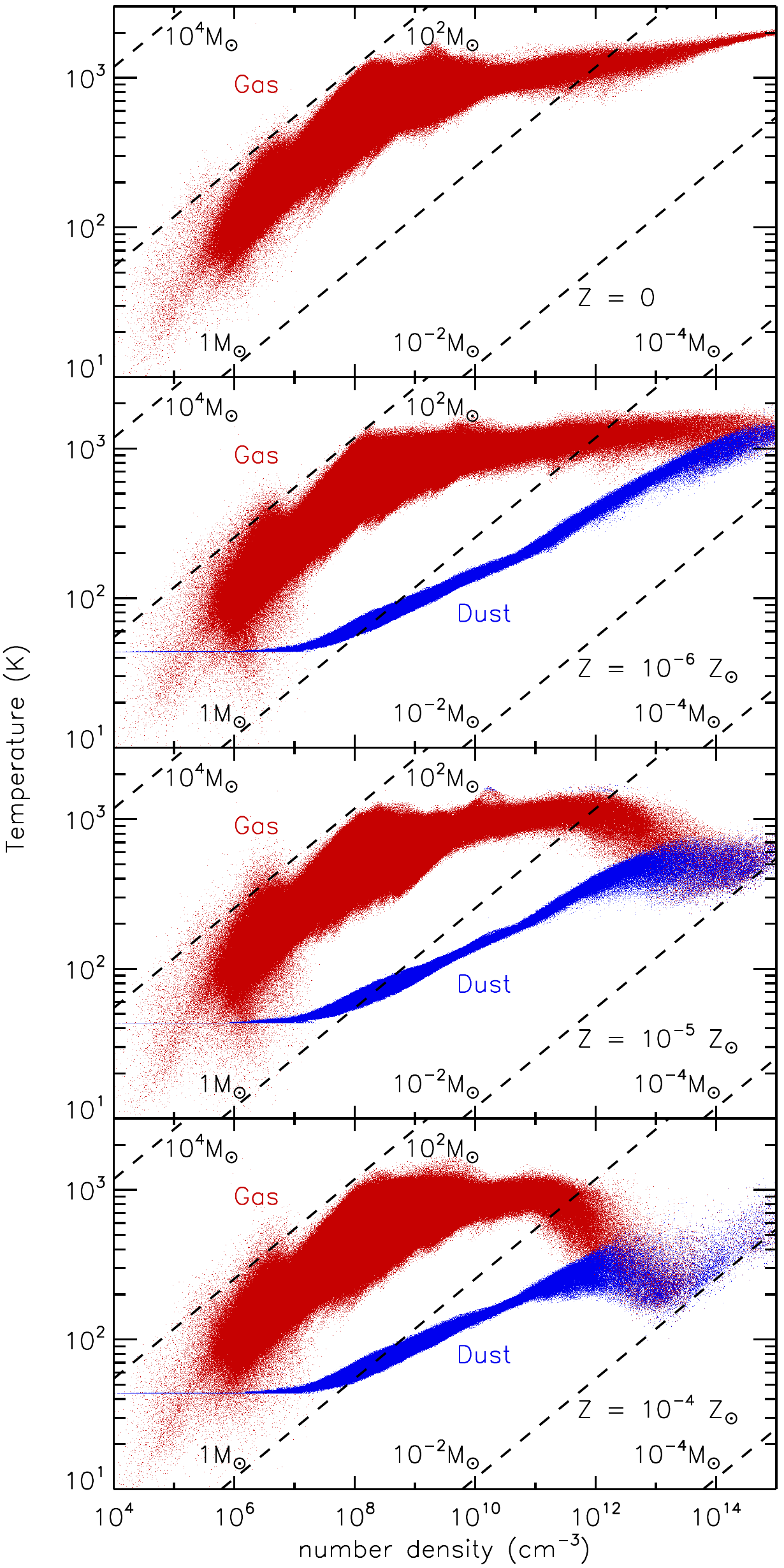}
  \caption{Dependence of gas and dust temperatures on gas density for metallicities $10^{-4}$, $10^{-5}$, and $10^{-6}$ and zero times the solar value, calculated just before the first sink particle was formed (see Table \ref{tsinks}). In red, we show the gas temperature, and in blue the dust temperature. The dashed lines are lines of constant Jeans mass.}  
 \label{nt}
\end{figure}
Collisions between gas particles and dust grains can transfer energy from the gas to the dust (if the gas temperature $T$ is
greater than the dust temperature $T_{\rm gr}$), or from the dust to the gas (if $T_{\rm gr} > T$). Full details of the dust cooling treatment can be found in \cite{2011ApJ...729L...3D}.

\subsection{Setup and Initial conditions}\label{setup}
We performed a set of four simulations, with metallicities Z/Z$_{\odot}$ = $10^{-4}$, $10^{-5}$, $10^{-6}$, and the metal-free case. Each simulation used 40 million SPH particles. We used these simulations to model the collapse of an initially uniform gas cloud with an initial number density of $10^{5} \: {\rm cm^{-3}}$ and an initial temperature of $300 \: {\rm K}$. The cloud mass was $1000 \: {\rm M_{\odot}}$. We included small amounts of turbulent and rotational energy, with $E_{\rm turb}/|E_{\rm grav}|  = 0.1$ and $\beta = E_{\rm rot}/|E_{\rm grav}| = 0.02$, where $E_{\rm grav}$ is the gravitational potential energy, $E_{\rm turb}$ is the turbulent kinetic energy, and $E_{\rm rot}$ is the rotational energy. The mass resolution is $2.5 \times 10^{-3} {\rm M}_{\odot}$, which corresponds to 100 times the SPH particle mass \citep[see e.g.][]{1997MNRAS.288.1060B}. The redshift chosen was $z = 15$, when the cosmic microwave background temperature was 43.6K. The dust properties were taken from \cite{2001ApJ...557..736G}, and the dust grain opacities were calculated in the same fashion as in \cite{2006MNRAS.373.1091B}. In the calculations, the opacities vary linearly with Z, which means for instance that for the Z/Z$_{\odot}$ = $10^{-4}$ calculations, the opacities were $10^{-4}$ times the original values.

\section{Analysis}\label{anal}
 
\subsection{Thermodynamical evolution of gas and dust}
Dust cooling is a consequence of inelastic gas-grain collisions, and thus the energy transfer from gas to dust vanishes when they have the same temperature. We therefore expect the cooling to cease when the dust reaches the gas temperature. In order to evaluate the effect of dust on the thermodynamic evolution of the gas and verify this assumption, we plot in Figure~\ref{nt}, the temperature and density for the various metallicities tested. We compare the evolution of the dust and gas temperatures in the simulations, at the point of time just before the formation of the first sink particle (see Table \ref{tsinks}). The dust temperature (shown in blue) varies from the CMB temperature in the low density region to the gas temperature (shown in red) at much higher densities.

Changes in metallicity influence the density at which dust cooling becomes efficient. For the Z $= 10^{-4}$ Z$_{\odot}$ case, dust cooling begins to be efficient at $n \approx 10^{11} \rm cm^{-3}$, while for Z $= 10^{-5}$ Z$_{\odot}$, the density where dust cooling becomes efficient increases to $n \approx 10^{13} \rm cm^{-3}$. For the Z $= 10^{-6}$ Z$_{\odot}$ case, dust cooling becomes important for $n \gtrsim 3 \times 10^{13}$ cm$^{-3}$, preventing the gas temperature from exceeding 1500~K. For comparison, in the metal-free case the gas reaches temperatures of approximately 2000~K.

The efficiency of the cooling is also expressed in the temperature drop at high densities.  The gas temperature decreases to roughly 400~K in the $10^{-5}$ Z$_{\odot}$ simulation, and 200~K in the Z $= 10^{-4}$ Z$_{\odot}$ case. This temperature drop significantly increases the number of Jeans masses present in the collapsing region, making the gas unstable to fragmentation.
The dust and the gas temperatures couple for high densities, when the compressional heating starts to dominate again over the dust cooling. The subsequent evolution of the gas is close to adiabatic.

When we compare our results to the calculations of \cite{2010ApJ...722.1793O}, we find good agreement with their 1D hydrodynamical models, although we expect some small difference due to effects of the turbulence and rotation \citep[see][]{2011ApJ...729L...3D} and also due to the use of different dust opacity models.

\subsection{Heating and cooling rates}
\begin{figure*}
\centering%
\includegraphics[width=1.0\linewidth,clip=]{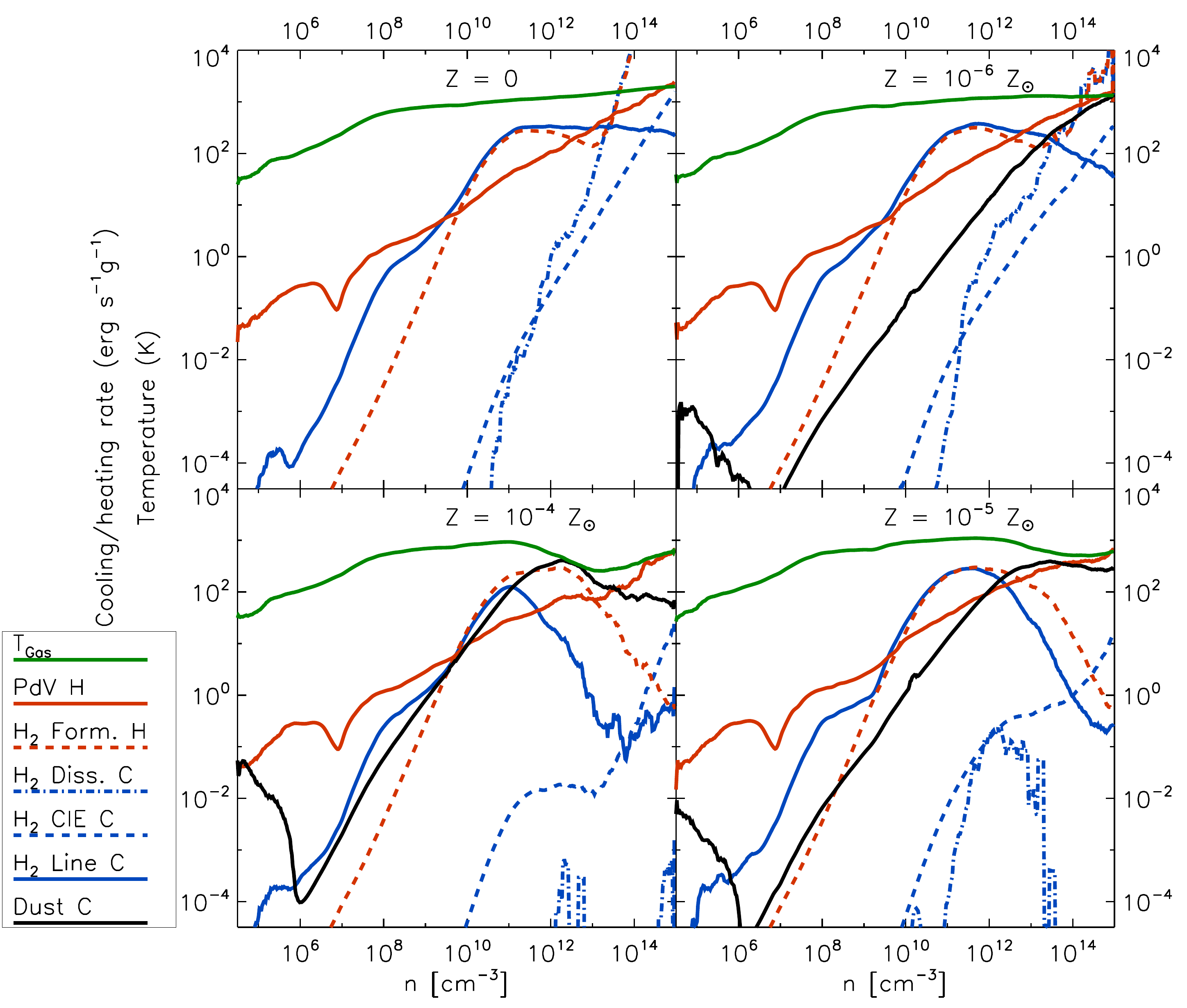}
\caption{Cooling and heating rates \emph{versus} number density for Z $= 10^{-4}$, 10$^{-5}$, 10$^{-6}$ Z$_{\odot}$, and zero. The values are calculated just before the first sink formed. The lines labeled as "C" indicate cooling, and "H" is heating. "Dust C", "H$_2$ Line C", "H$_2$ CIE" , and "H$_2$ Diss." indicate dust grain cooling, H$_2$ line emission, collision-induced emission, and dissociation cooling, respectively. "H$_2$ Form. H" and "$pdV$ H" are the H$_2$ formation heating rate, and compressive ($pdV$) heating rate.}
\label{ntc}
\end{figure*}

The thermal evolution of the gas during the collapse takes different paths depending on the metallicity, as shown in the density-temperature diagram (Figure \ref{nt}). In order to explain this behavior, we take a closer look at the cooling and heating processes involved. In Figure \ref{ntc} we show the main cooling and heating rates divided into four panels for the different metallicities.
These rates were calculated by averaging values of individual SPH particles in one density bin, where the total density range was divided in 500 bins in log space. 

At densities below $n \approx 10^{10}  \rm cm^{-3}$, dust cooling is unimportant in all of the runs. At these densities, the dominant coolant is H$_2$ line emission, while the heating is dominated by compressional ($pdV$) heating at $n \lesssim 10^{8}  \rm cm^{-3}$, and by three body H$_2$ formation heating at higher densities. 

At higher densities, dust cooling starts to play a more important role. 
In the Z $= 10^{-4}Z_{\odot}$ simulation, dust cooling exceeds $pdV$ heating at $n \approx 10^{10} \rm cm^{-3}$, although it does not exceed the H$_2$ formation heating rate until $n \approx 10^{11} \rm cm^{-3}$. Once this occurs, and dust cooling dominates, the gas temperature drops sharply. In the Z $= 10^{-5}Z_{\odot}$ simulation, on the other hand, dust becomes the dominant coolant only at $n \approx 10^{13} \rm cm^{-3}$, and so the temperature decrease happens later and is smaller. Finally, in the Z $= 10^{-6}Z_{\odot}$ case, dust cooling becomes competitive with $pdV$ heating only at the very end of the simulation, and so the effect on temperature evolution is less evident.

The other thermal processes play a minor role during the collapse. For example, H$_2$ dissociation cooling only becomes important in the runs with Z $= 10^{-6}Z_{\odot}$ and $0$, and only for $n > 10^{13} \rm cm^{-3}$. At very high densities ($n > 10^{14} \rm cm^{-3}$), H$_2$ collision-induced emission (CIE) cooling also begins to be important. For more details on H$_2$ heating and cooling processes in this very high density regime, we refer to \cite{2011ApJ...727..110C}.

\subsection{Fragmentation}\label{frag}
The transport of angular momentum to smaller scales during the collapse leads to the formation of a dense disk-like structure, supported by rotation. This disk then fragments into multiple objects.

\begin{figure}
  \centering
    \includegraphics[width=0.37\textwidth]{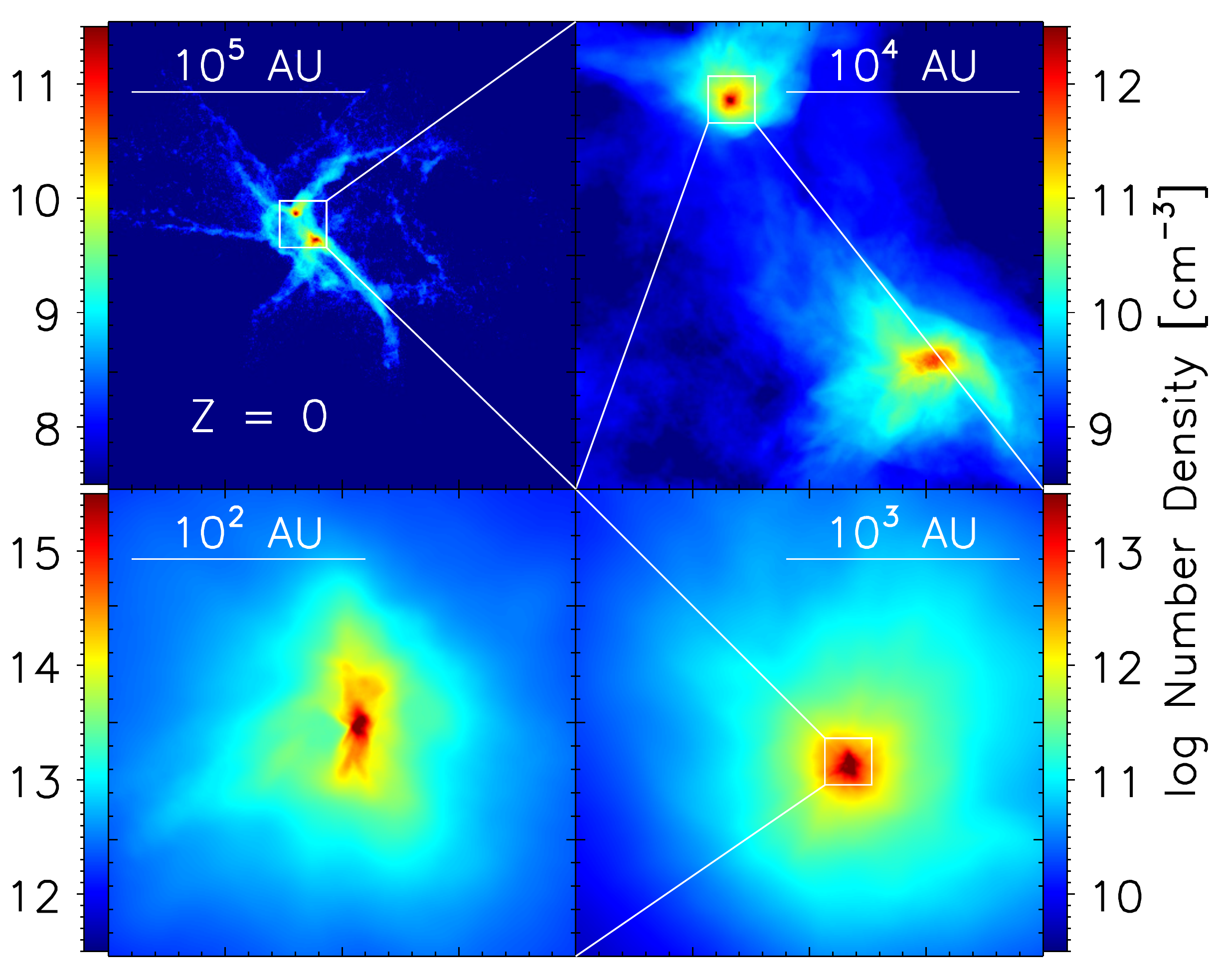}
    \includegraphics[width=0.37\textwidth]{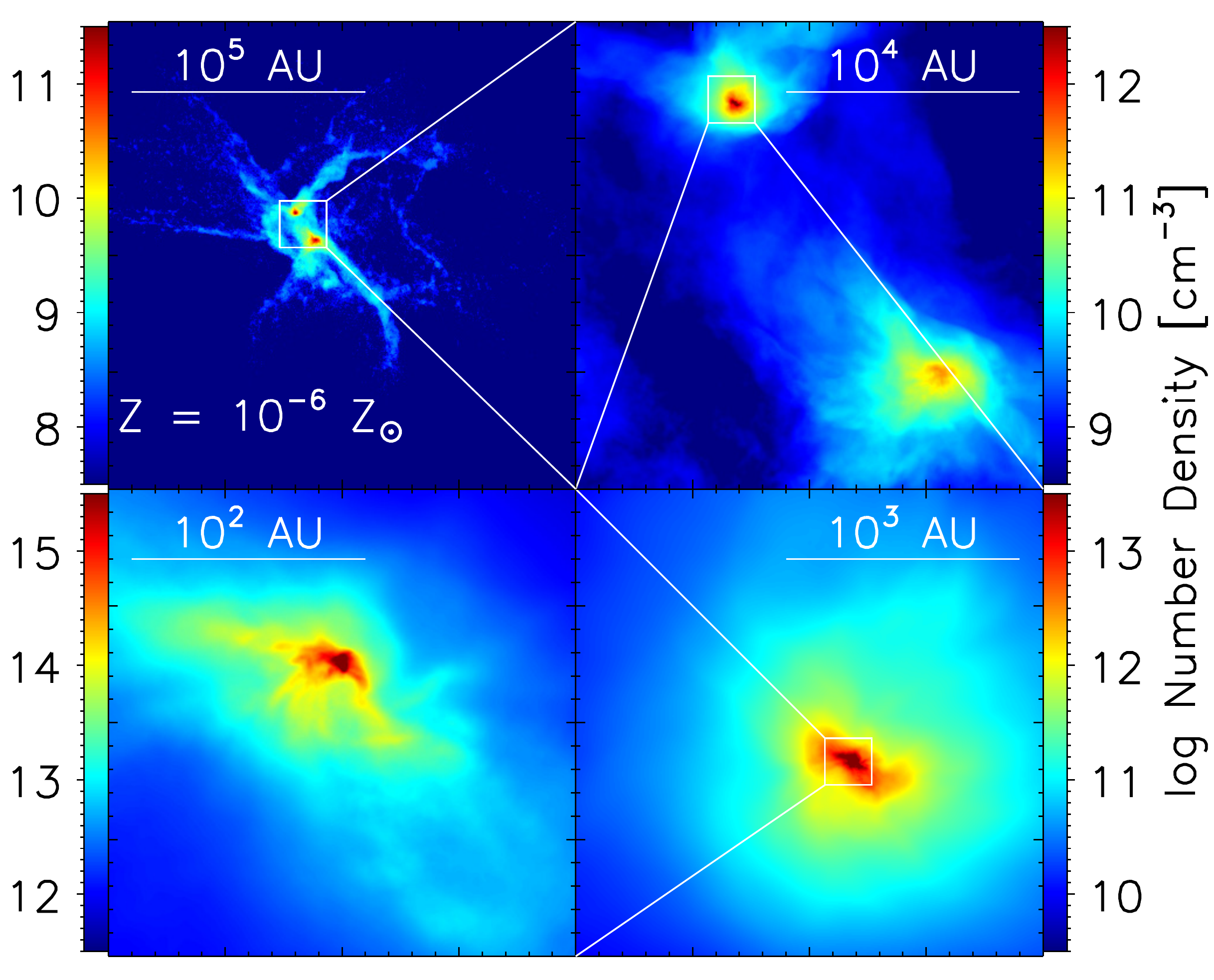}
    \includegraphics[width=0.37\textwidth]{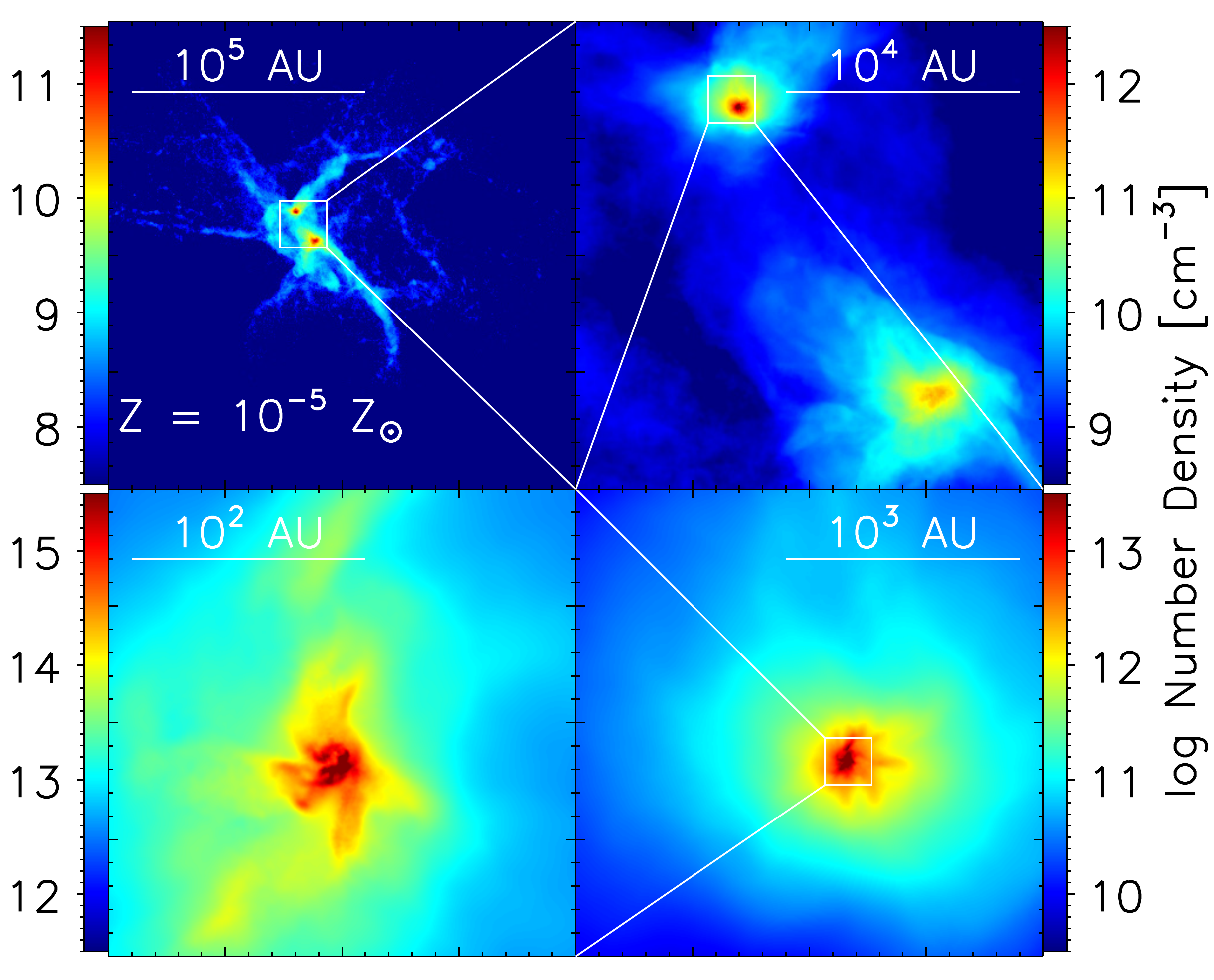}
    \includegraphics[width=0.37\textwidth]{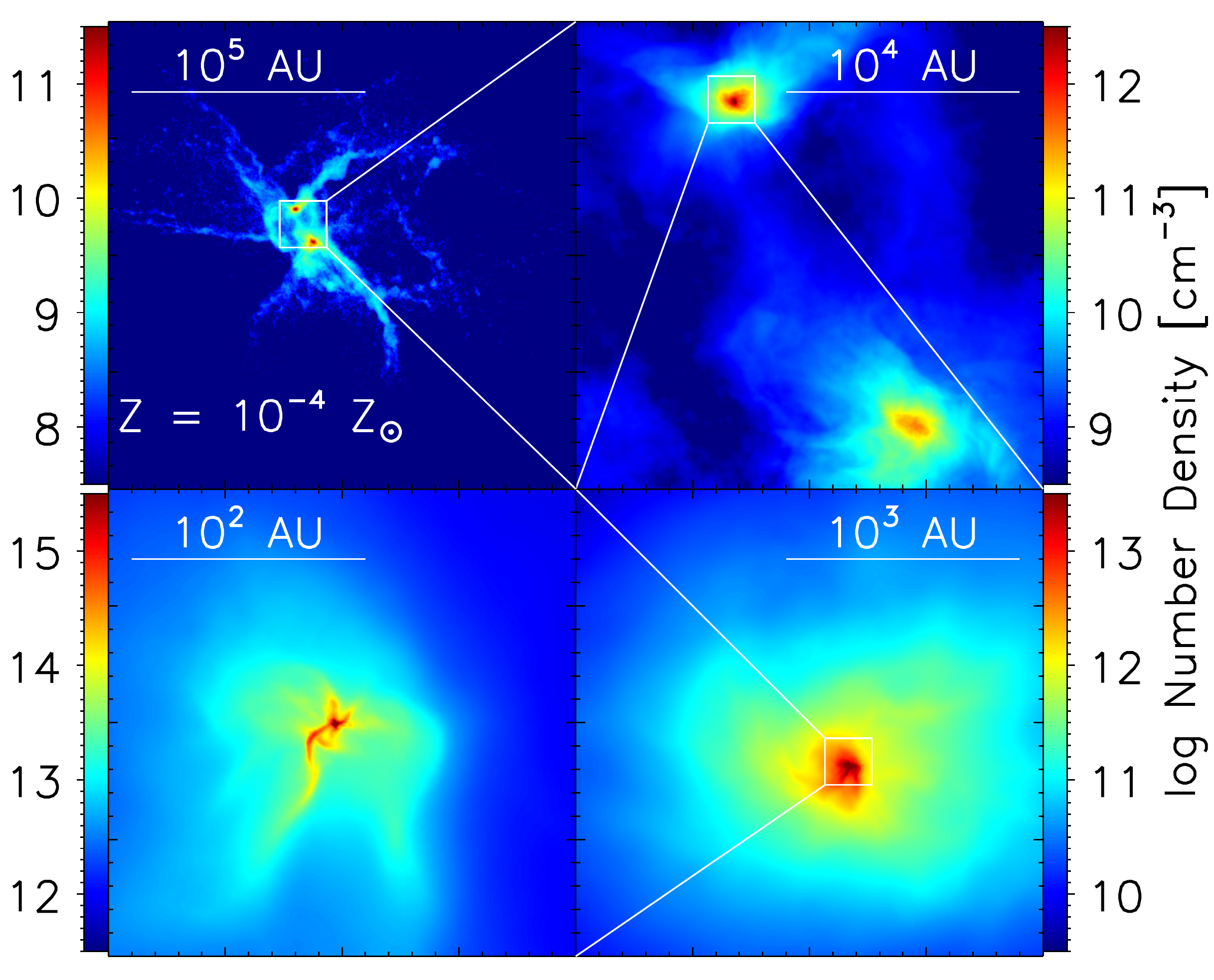}
    \caption{Number density maps for a slice through the high density region for Z $= 10^{-4}$ Z$_{\odot}$ (top), $10^{-5}$ Z$_{\odot}$, $10^{-6}$ Z$_{\odot}$, and 0 (bottom). The image shows a sequence of zooms in on the density structure in the gas immediately before the formation of the first protostar.}
    \label{nall}
\end{figure}
Figure~\ref{nall} shows the density structure of the gas immediately before the formation of the first protostar. The top-left panel shows a density slice on a scale comparable to the size of the initial gas distribution. The structure is very filamentary and there are two main over-dense clumps in the center. If we zoom in on one of the clumps, we see that its internal structure is also filamentary. Observe that at large scales the gas cloud properties are the same for all metallicities. Differences in the thermodynamic evolution appear only at $n \gtrsim 10^{11}$ cm$^{-3}$ (see Figure \ref{nt}). As a consequence, we observe variations in the cloud structure only in the high-density regions.

\begin{figure*}  
  \centering
    \includegraphics[width=0.9\textwidth]{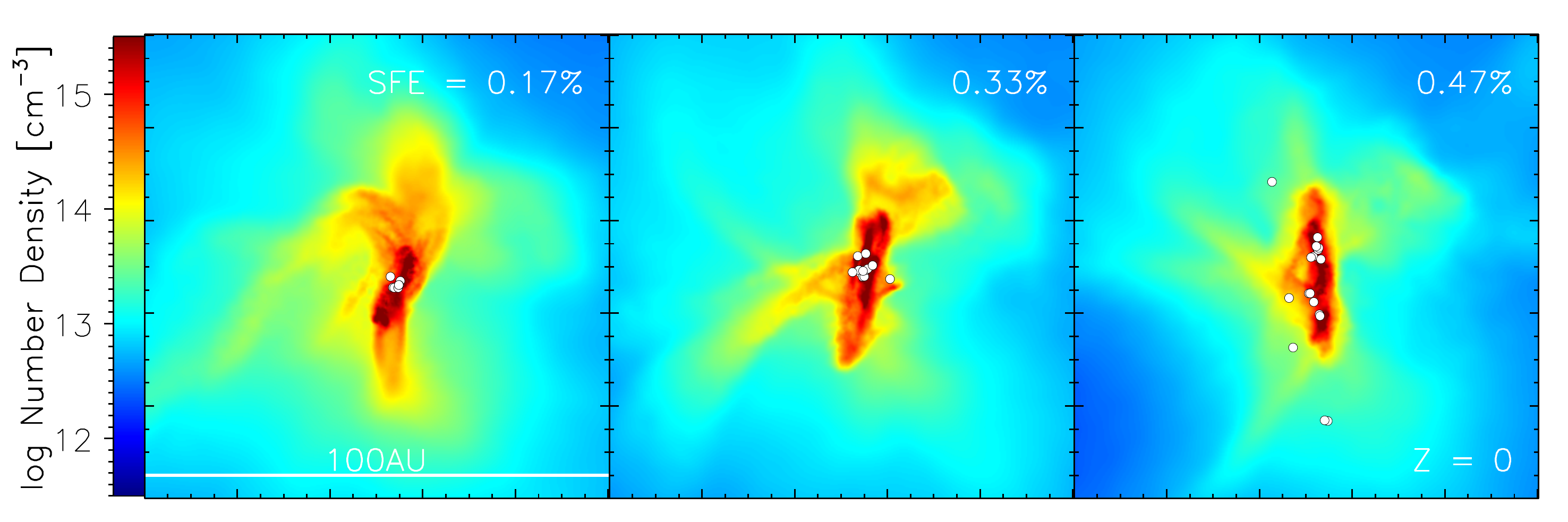}    
    \includegraphics[width=0.9\textwidth]{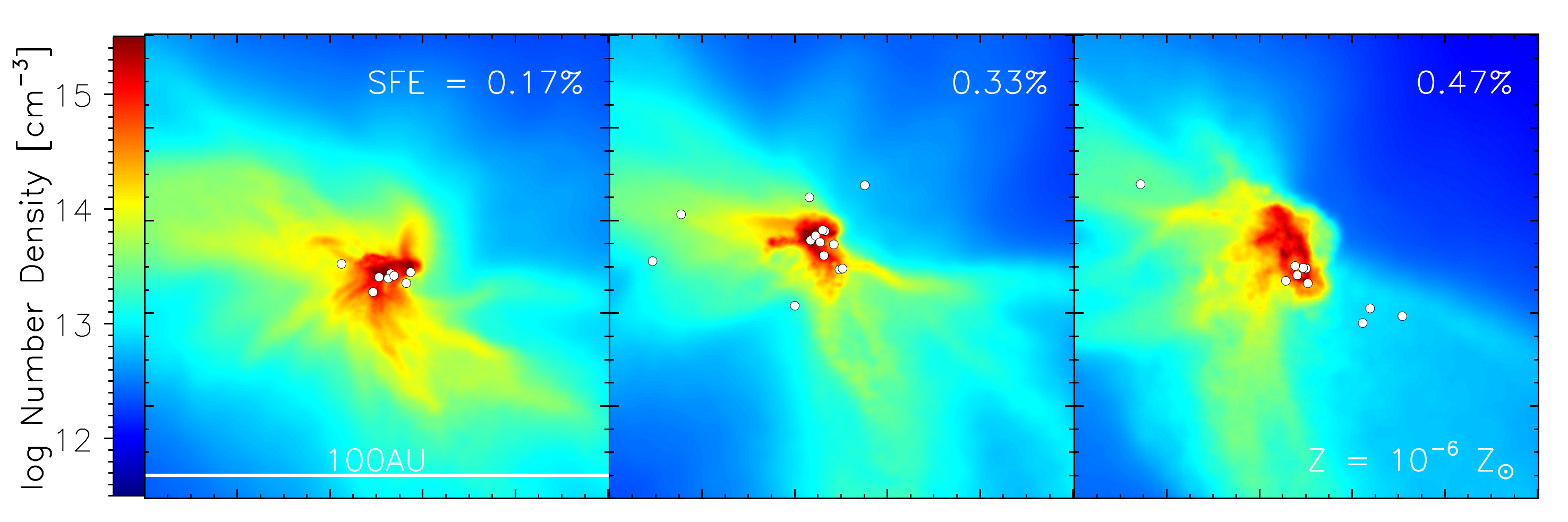}
    \includegraphics[width=0.9\textwidth]{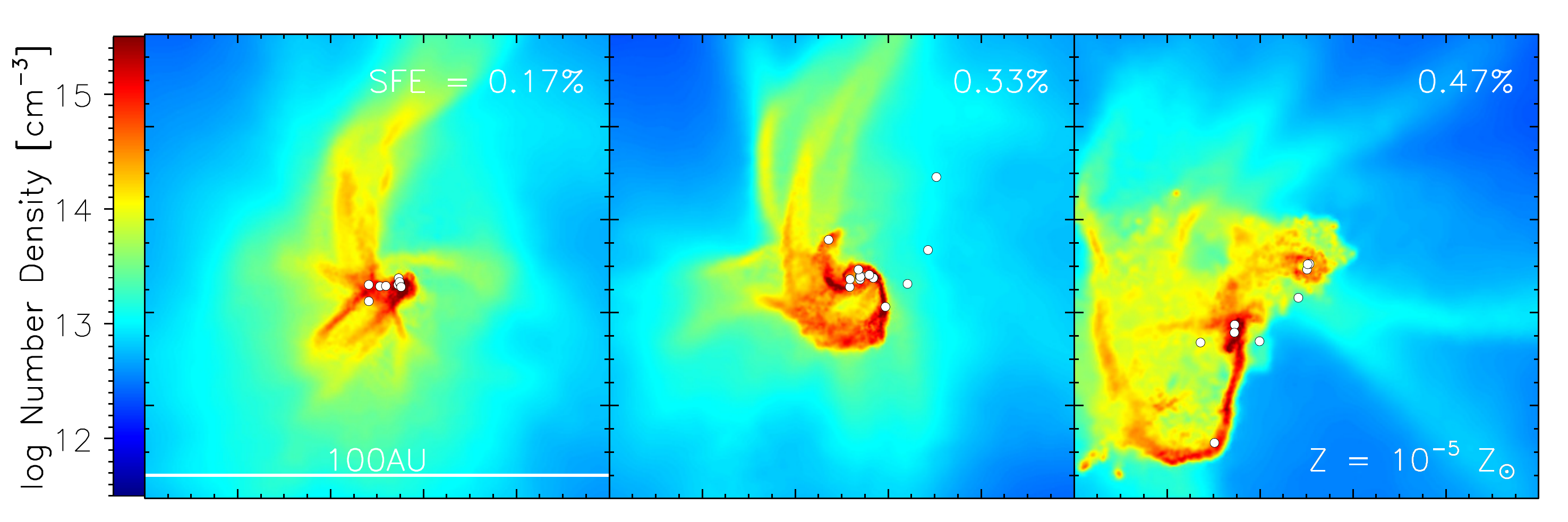}
    \includegraphics[width=0.9\textwidth]{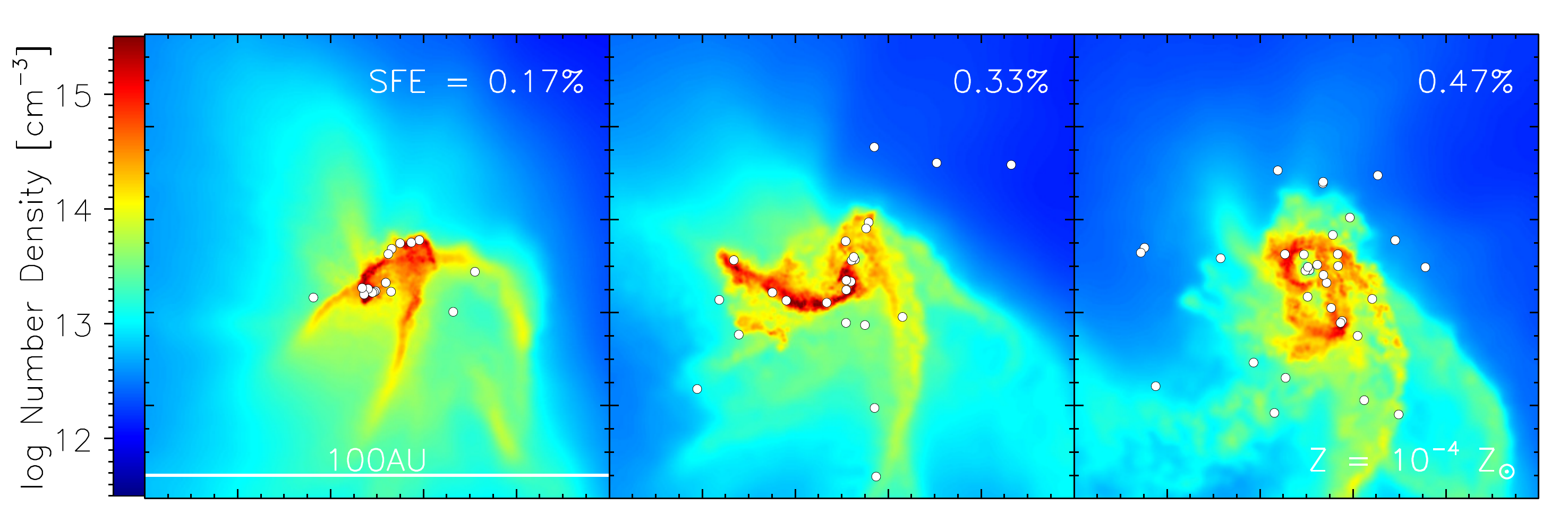}
    \caption{Number density map showing a slice through the densest clump, and the star formation efficiency (SFE) for Z $= 10^{-4}$ Z$_{\odot}$ (bottom), 10$^{-5}$ Z$_{\odot}$,  10$^{-6}$ Z$_{\odot}$, and 0 (top). The box is 100AU x 100AU and the percentage indicates the star formation efficiency, i.e. the total mass in the sinks divided by the cloud mass (1000M$_{\odot}$).\label{clumpall}}
\end{figure*}

Once the conditions for sink particle creation are met (see Section \ref{meth}), they start to form in the highest density regions (Figure \ref{clumpall}). Then, a disk is built up in these regions, where fragmentation also occurs \citep{1980ApJ...239..417T}. During further collapse, this dense region creates spiral structures.
For Z $= 10^{-5}$ Z$_{\odot}$ and $10^{-4}$ Z$_{\odot}$, density waves build up spiral structures, which become locally gravitationally unstable and go into collapse. The formation of binary systems by triple encounters \citep{2008gady.book.....B} transfers kinetic energy to some sink particles, causing them to be ejected from the high density region. For Z $= 10^{-5}$ Z$_{\odot}$, when the star formation efficiency (SFE) is 0.5\%, fragmentation has already occurred in a secondary dense center, at a distance of $\sim$ 20 AU from the first dense region.

For Z $= 10^{-6}$ Z$_{\odot}$ and 0, the formation of spiral structures is not observed. In these two runs, star formation occurs mainly in the central clump.

\begin{figure}
\centering
\includegraphics[width=1.0\linewidth]{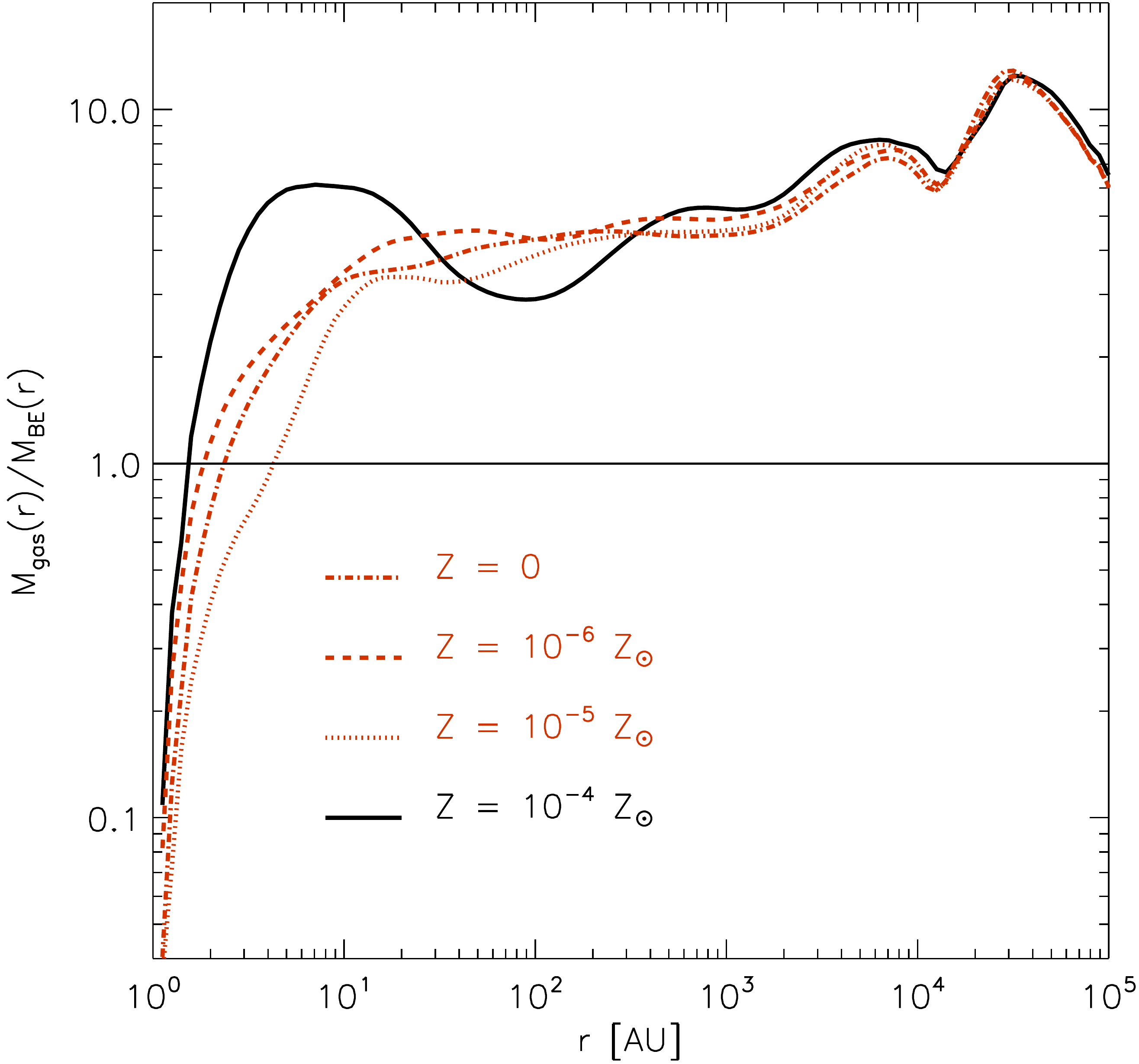}
\caption{Enclosed gas mass divided by Bonnor-Ebert mass \emph{versus} radius for different metallicities. The values were calculated at the time just before the first sink was formed and the center is taken to be the position of the densest SPH particle.}
\label{rjeans}
\end{figure}

One way to study the effect of dust cooling on the fragmentation behavior and the final stellar IMF is to look at the changes in the number of Bonnor-Ebert ($M_{BE}$) masses contained in this central dense region. Using the definition from \cite{2009Natur.459...49B}, 

\begin{equation}\label{bemass}
M_{BE} = 500M_{\odot} \left(\frac{T}{200 \rm K}\right)^{3/2} \left(\frac{n}{10^4 \rm cm^{-3}}\right)^{-1/2},
\end{equation}
for an atomic gas with temperature $T$ and number density $n$, we have computed the number of Bonnor-Ebert masses contained within a series of concentric radial spheres centered on the densest point in each of our four simulations. The results are shown in Figure \ref{rjeans}.

At the beginning of the simulation, the cloud had $\sim$ 3 $M_{BE}$. During the collapse, the gas cools and reaches $\sim$ 6 $M_{BE}$ in all cases. Cooling and heating are different depending on the metallicity, and this difference is seen for distances smaller than $\sim 400$ AU. The Z $= 10^{-4}$ Z$_{\odot}$ case, for instance, has twice the number of $M_{BE}$ for distances smaller than $\sim 10$ AU, when compared to the other cases. This will have direct consequences for the fragment mass function as we will see in the next section.

\subsection{Properties of the fragments}\label{fragprop}
\begin{figure}
  \centering
    \includegraphics[width=0.479\textwidth]{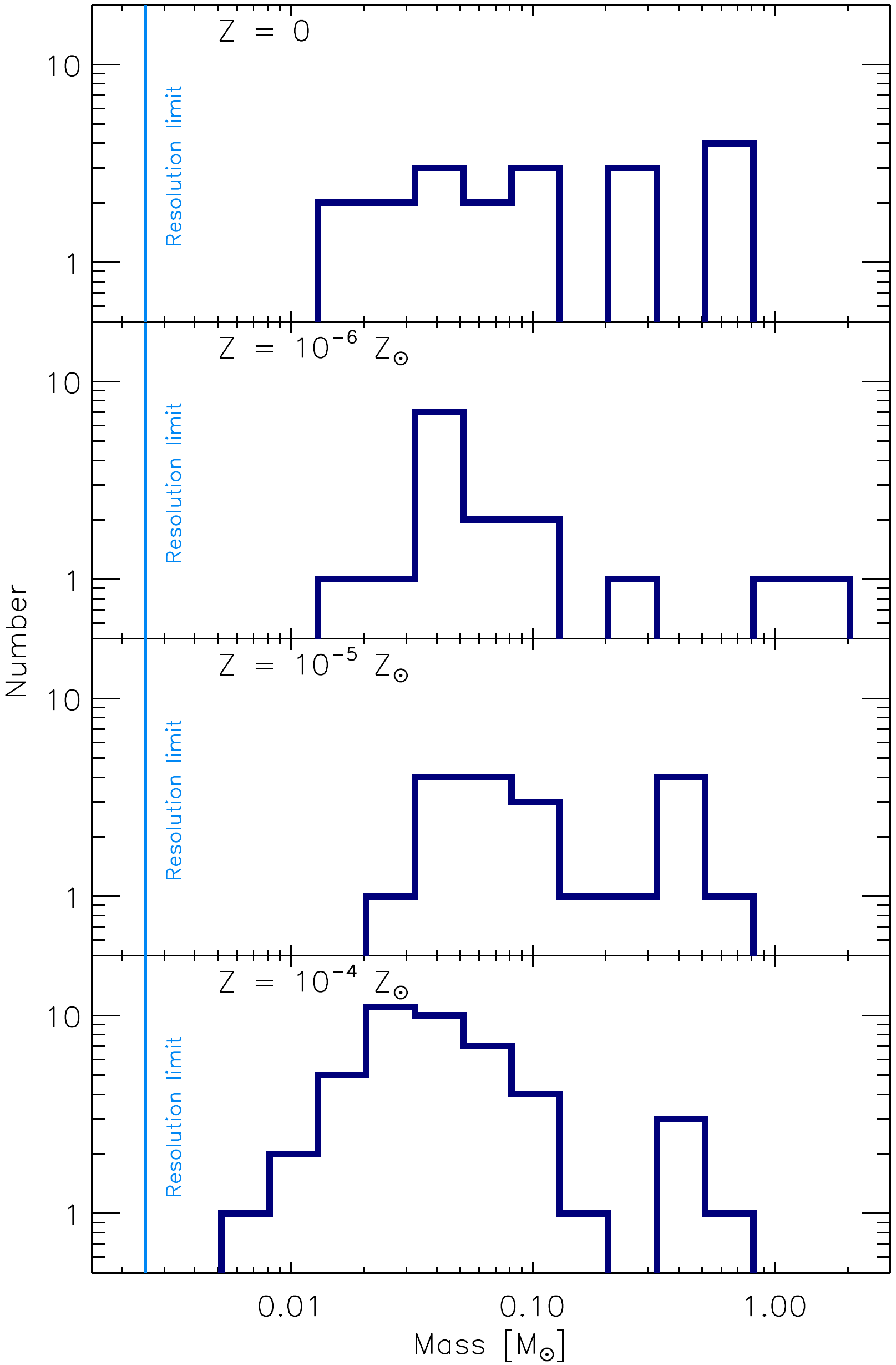}
    \caption{Sink particle mass function at the point when 4.7 $M_{\odot}$ of gas had been accreted by the sink particles. The mass resolution of the simulations is indicated by the vertical line.\label{sinkmf}}
\end{figure}

The simulations were stopped at a point when 4.7 $M_{\odot}$ of gas has been accreted into the sink particles, because the high computational cost made it impractical to continue. Figure \ref{sinkmf} shows the mass distribution of sink particles at that time. We typically find sink masses below 1 $M_{\odot}$, with somewhat smaller values in the $10^{-4}$ Z$_{\odot}$ case compared to the other cases. No sharp transition in fragmentation behavior was found, but rather a smooth and complex interaction between kinematic and thermodynamic properties of the cloud.

\begin{table}
\begin{center}
\begin{tabular}{ccrccccc}
\hline
Z/Z$_{\odot}$ & ST &  FT & SFR & Mean &Median&N\\
&(10$^3$yr)&(yr)&(M$_{\odot}$/yr)&(M$_{\odot}$)&(M$_{\odot}$)&\\
\hline\hline
            0    & 171.6 & 73   & 0.064& 0.24& 0.12 &19\\
10$^{-6}$ & 171.2 & 72   & 0.065& 0.29& 0.06 & 16\\
10$^{-5}$ & 170.8 & 88   & 0.053&0.24& 0.11 & 19\\
10$^{-4}$ & 169.2 & 138 & 0.034&0.10& 0.05 & 45\\
\hline
\end{tabular}
\end{center}
\caption{Sink particle properties for the different metallicities at the point where 4.7 $M_{\odot}$ have been accreted by the sink particles. "ST" (start time) is the time when sink particles start to form. "FT" (formation time), is the time taken to accrete 4.7 $M_{\odot}$ in the sinks. "SFR" is the mean star formation rate. Mean and median refer to the final mean and median sink mass. Finally, "N" is the number of sink particles formed.}
\label{tsinks}
\end{table}

Table \ref{tsinks} lists the main sink particle properties. It shows that the time taken to form the first sink particle is slightly shorter for higher metallicities. This shorter time is a consequence of the more efficient cooling by dust, which decreases the thermal energy that was delaying the gravitational collapse. In Table \ref{tsinks} we also observe that the star formation rate is lower for Z = $10^{-4}$ Z$_{\odot}$. This is because star formation started at an earlier stage of the collapse, when the mean density of the cloud was lower and there was less dense gas  available to form stars.

To better understand whether the resulting stellar cluster was affected by varying the metallicity, we plot the final sink mass distribution in Figure \ref{sinkmf}. It shows that for the simulations with $Z \leq 10^{-5}$ Z$_{\odot}$, the resulting sink particle mass function is relatively flat. There are roughly equal numbers of low-mass and high-mass stars, implying that most of the mass is to be found in the high-mass objects. This mass function is consistent with those found in other recent studies of fragmentation in metal-free gas \citep{2011ApJ...737...75G, 2011MNRAS.414.3633S}. If the sink particle mass function provides a reliable guide to the form of the final stellar IMF, it suggests that at these metallicities, the IMF will be dominated by high-mass stars.

All of the histograms in Figure \ref{sinkmf} have the lowest sink particle mass well above the resolution limit of $0.0025M_{\odot}$. Note that in all cases, we are still looking at the very early stages of star cluster evolution. As a consequence, the sink particle masses in Figure \ref{sinkmf} are not the same as the final protostellar masses --  there are many mechanisms that will affect the mass function, such as continuing accretion, mergers between the newly formed protostars, feedback from winds, jets and luminosity accretion, etc (see Section \ref{caveats}).

\subsection{Timescales}\label{sec:timescales}

\begin{figure}
\centering
\includegraphics[width=1.0\linewidth,clip=]{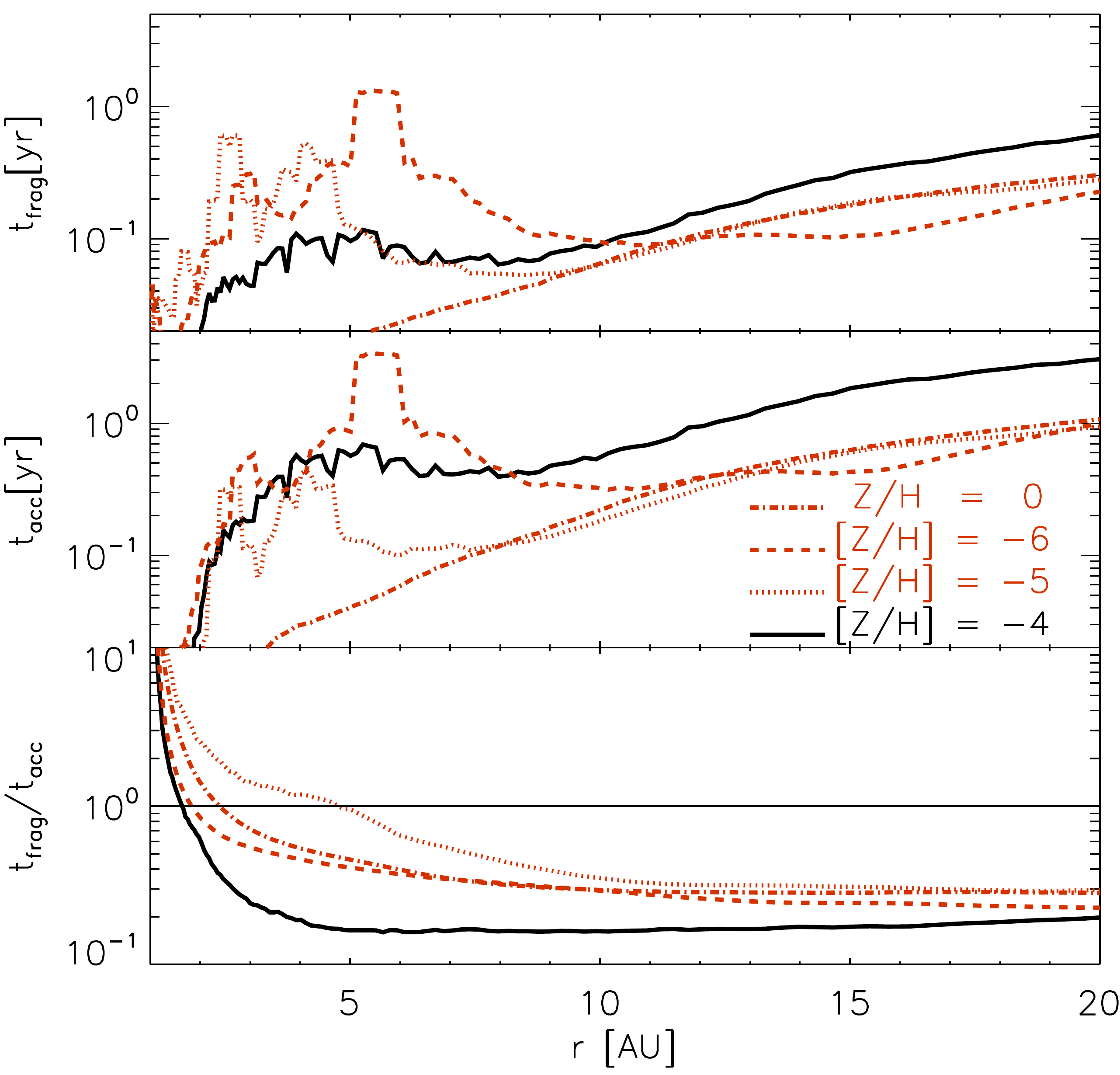}
\caption{Timescales for fragmentation (top panel) and accretion (middle panel), and also their ratio (bottom panel) according to the radius for the metallicities tested. The values were calculated just before the first sink particle was formed.}
\label{tfta}
\end{figure}

One way to explain the final mass distribution of the fragments is to look at the timescales for mass accretion and fragmentation. The degree of gravitational instability inside a volume can be represented by the number of Bonnor-Ebert masses contained in this volume. We can therefore estimate the fragmentation timescale by computing the time taken for the central dense region to accrete one Bonnor-Ebert mass. In other words, we have $t_{\rm frag} \equiv \rm M_{BE}/\dot{\rm M}$, where $\dot{\rm M}$ is the accretion rate. This value is shown as a function of the enclosed gas mass in Figure \ref{tfta}, where the values are calculated for particles in spherical shells, and the center is taken to be the densest SPH particle. $\dot{M}$ is obtained by summing up the mass of the particles ($m_p$) inside a shell times their radial velocity ($v_r$), $\dot{M} \equiv \displaystyle\sum\limits_{shell} m_{p}v_{r}$.

For comparison, we also plot the accretion timescale, here defined as the time taken by the gas to accrete the mass enclosed by that radius, $t_{\rm acc} \equiv \rm M_{\rm enc}/\dot{\rm M}$. When the fraction $t_{\rm frag}/t_{\rm acc} > 1$, one expects that the gas enclosed by this shell is going to be accreted faster than it can fragment, favoring high mass objects. Conversely, for $t_{\rm frag}/t_{\rm acc} < 1$, the gas will fragment faster than it can be accreted by the existing fragments, and the final mass distribution is expected to have more low mass objects. Note that as defined here, the timescale on which new fragments form $t_{\rm frag}/t_{\rm acc}$ is the inverse of the quantity M$_{\rm gas}/$M$_{\rm BE}$ plotted in Figure \ref{rjeans}.

In Figure \ref{tfta}, the simulation with Z $= 10^{-4}$ Z$_{\odot}$ has the lowest values for $t_{\rm frag}/t_{\rm acc}$, over a wide range of M$_{\rm enc}$. This indicates that more low-mass fragments are expected to form in this case, leading to a steeper fragment mass function.

\begin{figure}
\centering%
\includegraphics[width=1.0\linewidth,clip=]{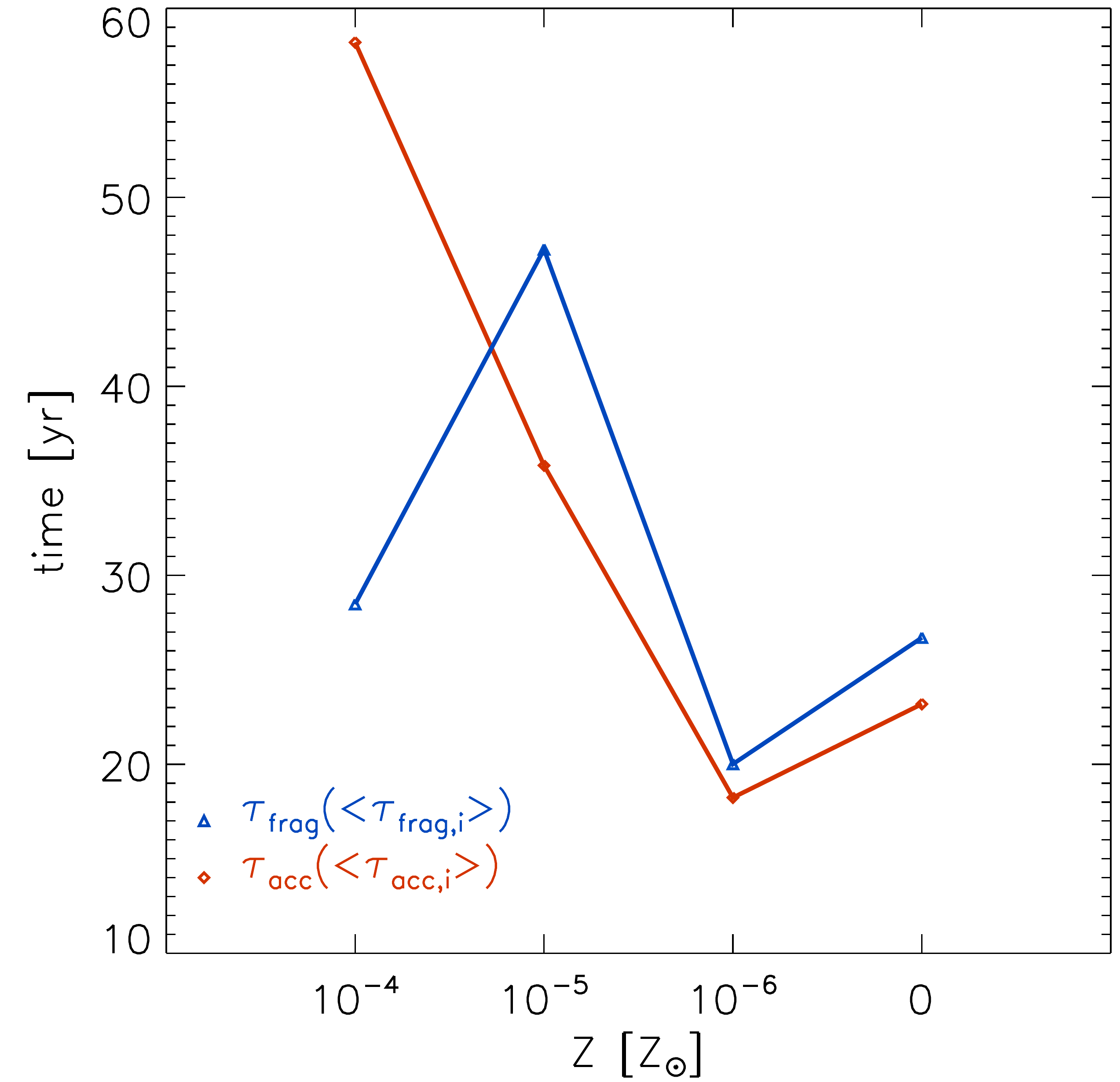}
\caption{Timescales for fragmentation and accretion for different metallicities calculated for the sink particles following Equations \ref{taufrag} and \ref{tauacc}.}
\label{timescales}
\end{figure}

Now we can compare the predicted values before sink formation started with the final accretion and fragmentation timescales. These values are designed to represent the characteristic timescales on which the mass histogram changes: the fragmentation timescale ($\tau_{\rm frag}$) is the time on which the number of fragments change by a significant amount, while the accretion timescale ($\tau_{\rm acc}$) represents the time on which the existing fragments grow in mass. We therefore define $\tau_{\rm frag} \equiv  n/\dot{n}$, and $\tau_{\rm acc} \equiv M/\dot{M}$, where $n$ is the number of sink particles, and M is the total mass incorporated into sink particles.

Both timescales should increase over time, since the number of sink particles and the total mass also increase. This reflects the fact that it takes longer to change the shape of the mass histogram in frequency when the number of elements is higher, and in mass when the total mass is high. The key point is that these timescales are related to the shape of the histogram. If the timescale for accretion is shorter than the timescale for fragmentation, the histogram will tend to be dominated by high-mass objects. Conversely, if the fragmentation timescale is shorter than the accretion timescale, the low-mass part will be populated before the objects can grow and occupy higher mass bins. Therefore, a comparison between $\tau_{\rm acc}$ and $\tau_{\rm frag}$ helps us to understand the shape of the mass histogram.

We calculate the average $\tau_{\rm acc}$ and $\tau_{\rm frag}$ at each point in time when a new sink particle is created, since $\tau_{\rm frag}$ is not defined for $\dot{n} = 0$.  The value for the mean $\tau_{\rm frag}$ and $\tau_{\rm acc}$ is calculated by averaging their individual times in equations \ref{taufrag} and \ref{tauacc}:

\begin{align}
\tau_{\rm frag,i} \equiv& \frac{n_i}{\dot{n_i}}   =  \frac{n_i}{\Delta n_i/\Delta t_i}  =  \frac{n_i(t_i - t_{i-1})}{n_i - n_{i-1}} ,\label{taufrag}\\
\tau_{\rm acc,i}  \equiv& \frac{M_i}{\dot{M_i}} = \frac{M_i}{\Delta M_i/\Delta t_i} = \frac{M_i (t_i - t_{i-1})}{M_i - M_{i-1}}\label{tauacc},
\end{align}
where $i$ is the point in time when a new sink is created, and it varies from 2 to the total number of sink particles. $n_i$, $M_i$, $t_i$ are the number of sink particles, the total mass in sinks, and the time at the point $i$.

Figure \ref{timescales} shows the average timescales for fragmentation and accretion for different metallicities. These results explain the difference in the sink particle mass distribution in Figure \ref{sinkmf}. For Z $\leq 10^{-5}$ Z$_{\odot}$, the fragmentation time is always larger than the accretion time, indicating that the sink particles will accrete faster than they can be generated, resulting in a flatter mass distribution. On the other hand, when the fragmentation time is longer than the accretion time (for Z $= 10^{-4}$ Z$_{\odot}$), the gas fragments, rather than moving to the center and being accreted. As a consequence, the low-mass end of the protostellar mass function grows faster than the high-mass end, and the slope of the mass function steepens. This behavior agrees well with the predictions from before fragmentation started, shown in Figure \ref{tfta}. 

Note that the values in Figure \ref{tfta} were calculated before the formation of the first sink particle, while the values in Figure \ref{timescales} were calculated using the sink particle properties. A comparison between them is useful to evaluate whether the gas cloud properties from before star formation started could be used to predict its star formation behavior. The trend in both figures is that the fragmentation timescale is normally shorter than the accretion timescales for Z $= 10^{-4}$ Z$_{\odot}$. From these results we conclude that the mass distribution in Figure \ref{sinkmf} can be explained by the timescales in Figure \ref{timescales}, in particular the fact that the Z $\leq 10^{-5}$ Z$_{\odot}$ simulations have more high-mass objects. The last finding is that the transition from Pop. III to Pop. II star formation mode is not abrupt, in the sense that there is no metallicity bellow which the gas cannot fragment. The transition is rather in the stellar initial mass function, and gas clouds with Z $\lesssim 10^{-5}$ Z$_{\odot}$ form a more flat IMF, while gas clouds with Z $\gtrsim 10^{-4}$ Z$_{\odot}$ produce a cluster with more low-mass objects \citep[see also][]{2008ApJ...672..757C}.

\subsection{Radial mass distribution}

\begin{figure}
\centering%
\includegraphics[width=1.0\linewidth,clip=]{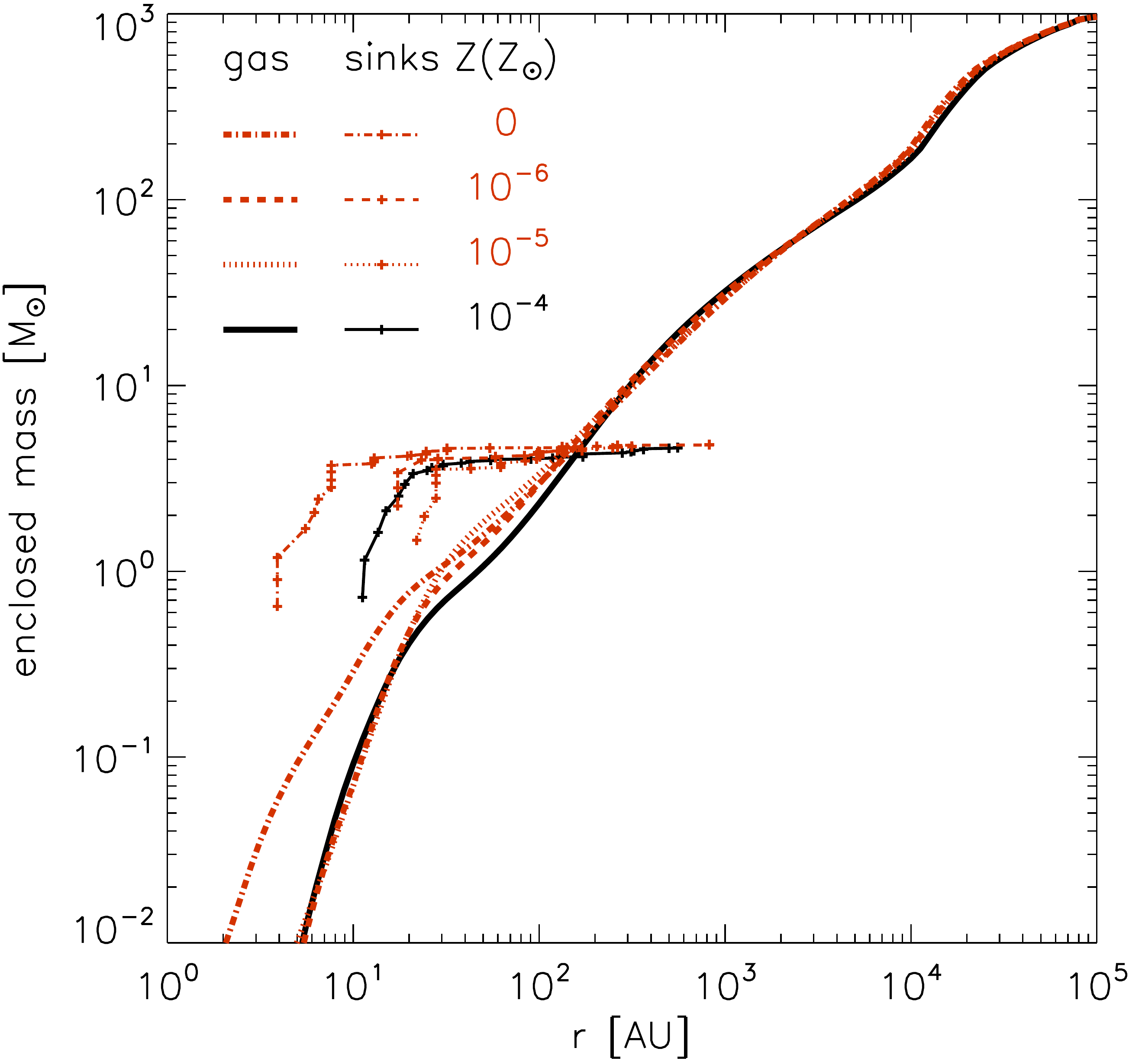}
\caption{Dependence of the enclosed gas and sink mass on the distance from the center of mass, for the four simulations. The values were calculated at a point when 4.7 $M_{\odot}$ of gas had been accreted.}
\label{rmegs}
\end{figure}
Another property of the star-forming cloud that we observed to vary in our calculations is the spatial mass distribution. The dependence of the enclosed gas and sink mass on the distance from the center of mass is shown in Figure \ref{rmegs}. The Z $= 0$ case has almost all of the sink particle mass concentrated within $r <$ 8AU. The gas density for this case is also higher in this region, when compared to the other metallicities, showing that the gas and sink particle mass densities follow each other. The mass in sink particles exceeds the gas mass for small radii, being the most important component in the gravitational potential. For $r > 150$ AU, the gas becomes the most massive component, for all Z. \cite{2012MNRAS.420..613G} also reported this behavior, but in their case the sink particles already started to dominate the potential below $r \approx 10^3$ AU. 

This higher concentration of gas and sinks at the center occurs because for the Z = 0 case, the gas had higher temperatures in the central region. For high temperatures, the criterion for gravitational instability requires higher densities, which are achieved only very close to the center. As a consequence, the sink particle formation criteria are met just for short distances from the center.

Consequently, the dominance of sink particles mass in the gravitational potential over the gas mass, for radii smaller than 150 AU, shows the importance of treating gas and stars together in this sort of problem. It also suggests that N-body effects, such as ejections and close encounters, should play an important role in the formation of these dense star clusters, even in the very earliest stages of their evolution \citep[see][]{2011MNRAS.414.3633S, 2012arXiv1202.5552G}.

\section{Caveats}\label{caveats}

Our aim with these calculations is to study the importance of dust cooling for fragmentation in high-redshift halos. To better understand star formation in this environment, additional physical processes should be considered as well.

Particularly, the low number of sink particles ($\approx 20$) and the small SFE (0.5\%) do not permit to constrain the stellar IMF adequately. By running the calculations until the SFE goes towards higher values, uncertainties involving the fragments that formed during the simulations can be diminished. However, this does not appear to be computationally feasible with our current approach.

The mass accreted by the sink particles varies with the different metallicities, and affects the final sink particle mass function. It also influences the expected accretion luminosity. We did not take this process into account during our calculations, but we can estimate its importance relative to other thermal processes.

\begin{figure}[!ht]
\centering
\includegraphics[width=1.0\linewidth]{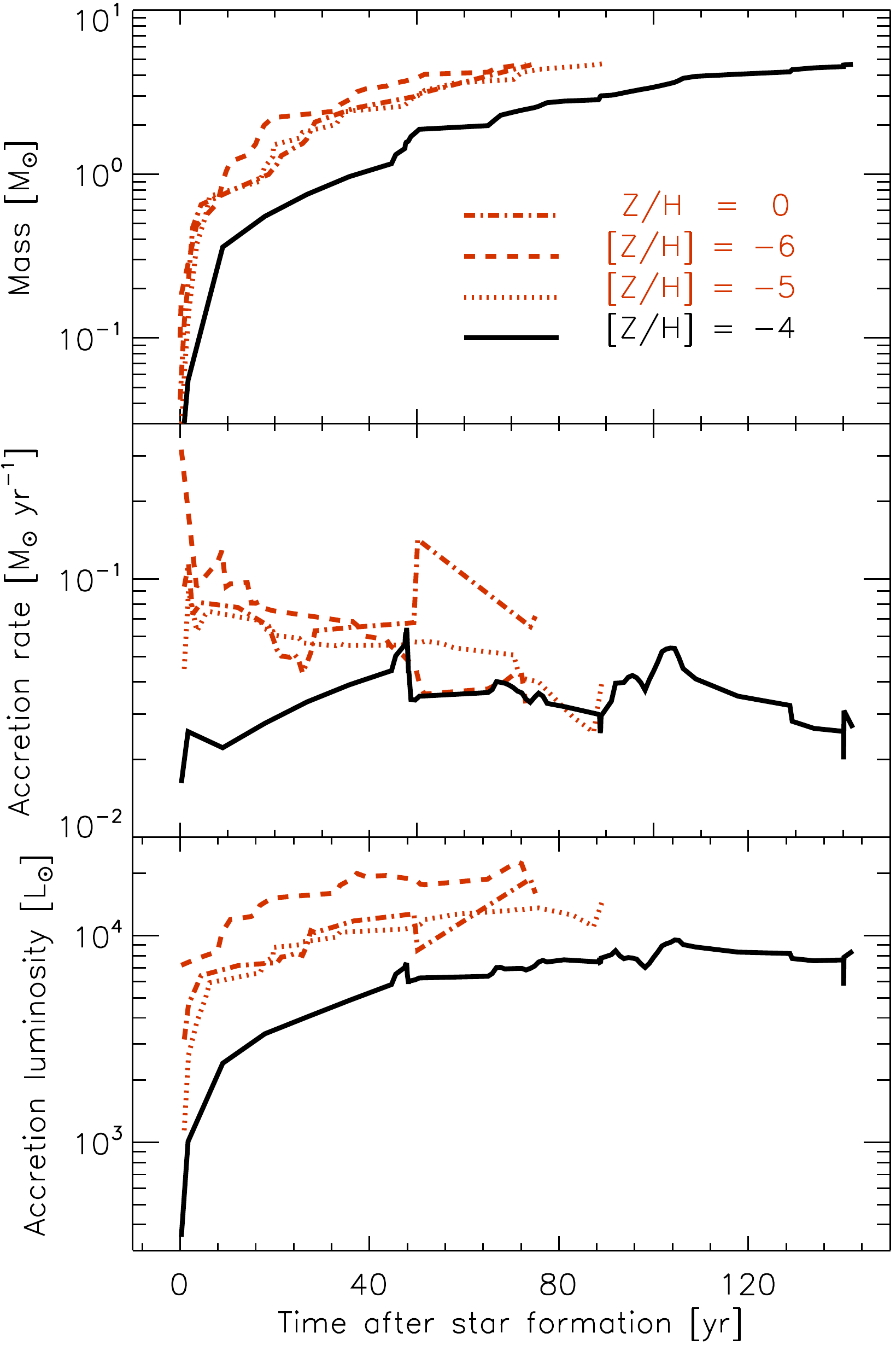}
\caption{Time evolution of the mass, mass accretion rate, and accretion luminosity for the four metallicities, for all sink particles combined.}
\label{mmdot}
\end{figure}

In Figure \ref{mmdot} we present the accretion properties for the newborn stellar systems. The top panel shows how the total mass in sinks evolve with time, for different metallicities. The accretion rate varies from 0.02 to 0.17 $M_{\odot}$ yr$^{-1}$, and it is on average lower for the Z $= 10^{-4}$ Z$_{\odot}$ case.

In the bottom panel of Figure \ref{mmdot}, we show the accretion luminosity calculated by adding up all sink particle contributions, with the standard equation,

\begin{equation}\label{acclum}
L_{acc} = \frac{G M_* \dot{M}}{R_*},
\end{equation}
where $\dot{M}$ is the mass accretion rate by a protostar with mass $M_*$, and stellar radius $R_*$. We calculate $R_*$ following \cite{1986ApJ...302..590S} using 
\begin{equation}
R_* = 66.8 \left(\frac{M_*}{M_{\odot}}\right)^{0.27} \left(\frac{\dot{M}}{10^{-2} M_{\odot} \rm yr^{-1}}\right)^{0.41} R_{\odot}.
\label{eqnsr}
\end{equation} 

The accretion luminosity varies from few times $10^3$ L$_{\odot}$ to around $50 \times 10^3$ L$_{\odot}$, depending on metallicity. For Z $\leq 10^{-5}$ Z$_{\odot}$, the accretion luminosity is always over $10^4$ L$_{\odot}$. The Z $= 10^{-4}$ Z$_{\odot}$ case has the lowest estimated accretion luminosity, around four times lower than the other cases. The values found for the accretion luminosity are similar to the ones found by \cite{2011MNRAS.414.3633S} for Z = 0, where they argue that accretion luminosity could delay the fragmentation, but not prevent it.

We can now compare this expected luminosity, and its consequent heating, to the heating processes in Figure \ref{ntc}. We make the assumption that gas and dust are absorbing the radiation in the optically thin regime. This overestimates the effects, and we obtain an upper limit for the accretion luminosity heating,
  \begin{equation}\label{ah}
  \begin{array}{rl}
   \Gamma_{acc} = \kappa_P \left(\dfrac{L_{acc}}{4 \pi r^2}\right) \rm erg ~s^{-1} g^{-1}
  \end{array}
\end{equation}
where $\rho_g$ is the gas density, $\kappa_P$ is the Planck mean opacity, and $r$ is the distance from the source. We also assume that heating occurs at $\rho \approx 10^{-9} \rm g~cm^{-3}$. With this assumption we can calculate the mean gas temperature, and use the Planck mean opacity for the gas from \cite{2005MNRAS.358..614M}, for their fiducial Pop.\,III chemical composition. The Planck mean opacity for the dust is calculated in the same way as in \cite{2011ApJ...729L...3D}. Finally, the combined Planck mean opacity is the sum of gas and dust contributions, $\kappa_P \equiv \kappa_{gas} + \kappa_{dust}$.

By considering the maximum accretion luminosity for each case, we get that $\Gamma_{acc} = 4.9$, 0.9, 1.7, and
$0.7 \times 10^{3}( \rm 20AU/r)^2  \rm erg~ s^{-1} g^{-1}$,
for $Z = 10^{-4}$, $10^{-5}$, $10^{-6}$, and 0 $Z_{\odot}$. 

As these values are comparable to the other thermal processes at high densities (see Figure \ref{ntc}), it would seem that accretion luminosity heating from the young protostars may have some effect on the way in the which the gas behaves. However without doing the radiative transfer explicitly, it is difficult to estimate how big this effect will be.

Although the amount of heating seems high, the dust cooling is a strong function of temperature in this regime, and so it could be that dust temperatures remain quite similar. One must also remember that the above estimates do not take into account the extinction and reprocessing of the radiation field that will occur in the optically thick region that surrounds the protostar. However even a factor of 2 change in the dust temperature will remove the dip in the $\rho - T$ phase diagrams that we show in Figure \ref{nt}, and thus remove the ability of the dust to set a new length-scale for fragmentation.

Sufficiently far enough away from the strongest sources, the effect will obviously drop to the point at which the physics in our current calculations are applicable. However if we look to Figure \ref{clumpall}, we see that most of the fragmentation that we report is confined to a few tens of AU around the central protostar, and as such, the effects of the accretion luminosity are likely to change the picture that we present in this paper to some extent. We hope to explore this effect in a future study.

Another aspect of our model that could be improved upon are the dust opacities. The thermal evolution can be calculated more accurately if we use dust opacities that better represent the values expected for very low metallicity environments. The dust opacity in our simulations correspond to values calculated for the Milky Way and then scaled with metallicity. This means that the opacity values for the Z = 10$^{-4}$Z$_{\odot}$ case are 10$^{-4}$ times the dust opacity in the Milky Way. This approximation is probably not fully correct, and the use of a more accurate model \citep[e.g.][]{2001MNRAS.325..726T, 2007MNRAS.378..973B, 2012MNRAS.419.1566S, 2003ApJ...598..785N, 2006ApJ...648..435N} can change the value of the cooling in the region where fragmentation occurs. This change affects the local Jeans mass, and consequently the star formation behavior.

Furthermore, the available models give the dust composition for different scenarios in the early Universe, e.g.\,different supernovae progenitor masses \citep{2012MNRAS.419.1566S}, and the use of such models would add another variable to the problem - the stellar population for the supernovae progenitors. One reasonable approach is to test different scenarios and see how they would affect the properties of the cluster of stars that forms. In this sense, the dust composition is a problem in itself that should be addressed. Since cooling affects the fragmentation behavior and mass accretion, a more realistic dust model improves the accuracy with which we can model star formation at low metallicities. We intend to address this issue in a future paper.

Another effect that could change our results is the possibility of inelastic encounters between the protostars. Star formation in our simulation occurs at very high densities, where inelastic encounters between the new born protostars could occur. In similar conditions to the ones tested here, \cite{2011MNRAS.414.3633S, 2011arXiv1112.4157S} show that the estimated stellar radius could be as large as $\sim$ 1 AU, a value comparable to the distances between the sink particles shown in Figure \ref{clumpall}. By not accounting for merging of such objects, we could be overestimating the final number of fragments, although we expect new protostars to continue to form. For a detailed investigation of this process see \cite{2012arXiv1202.5552G}.

Finally, the inclusion of magnetic fields in the calculations could alter the fragmentation picture as it is presented in this study. They can be amplified during gravitational collapse \citep{2010A&A...522A.115S}, generating values strong enough to delay the collapse \citep{2009ApJ...703.1096S, 2010ApJ...721L.134S, 2011ApJ...731...62F, 2011PhRvL.107k4504F, 2012ApJ...745..154T}. Analytic amplification values are calculated by \cite{2012PhRvE..85b6303S}. From modeling present-day star formation, we know that the presence of magnetic fields can decrease the level of fragmentation, but cannot prevent it, for the expected saturation levels of a few percent \citep{2011ApJ...729...72P, 2011A&A...528A..72H}. 

\section{Conclusions}\label{conc}
In this paper we have addressed the question of whether dust cooling can lead to the fragmentation of low-metallicity star-forming clouds. For this purpose we performed numerical simulations that follow the thermodynamical and chemical evolution of collapsing clouds. The chemical model included a primordial chemical network together with a description of dust effects, where the dust temperature was calculated by solving self-consistently the thermal energy equilibrium equation.

As a result, we found that dust can cool the gas, for number densities higher than $10^{11}$, $10^{12}$, and $3 \times 10^{13} \rm cm^{-3}$ for Z $= 10^{-4}$, $10^{-5}$, and $10^{-6}$ Z$_{\odot}$, respectively. Higher metallicity implies larger dust-to-gas fraction, and consequently stronger cooling. This is reflected in a lower temperature of the dense gas for the higher metallicity simulations, and this colder gas permitted a faster collapse. Therefore, the fragmentation behavior of the gas depends on the metallicity, and higher metallicities lead to a faster collapse.

For example, the characteristic fragment mass was lower for $Z =  10^{-4}$ Z$_{\odot}$, since a lower temperature reduces the Bonnor-Ebert masses at the point where the gas undergoes fragmentation. This also implies a lower ratio of fragmentation and accretion time, $t_{\rm frag}/t_{\rm acc}$, which will lead to a mass function dominated by low-mass objects. For Z $\leq  10^{-5}$ Z$_{\odot}$, fragmentation and accretion timescales are comparable, and the resulting mass spectrum is rather flat, with roughly equal numbers of stars in each mass bin.

In addition to that, dust cooling appears to be insufficient to change the stellar mass distribution for the Z $= 10^{-5}$ and $10^{-6}$ Z$_{\odot}$ cases, when compared with the metal-free case. This can be seen in the sink particle mass function (Figure \ref{sinkmf}), which shows that the Z $\leq 10^{-5}$ Z$_{\odot}$ cases do not appear to be fundamentally different.


Finally, we conclude that the dust is not an efficient coolant at metallicities below or equal to Z$_{\rm crit} = 10^{-5}$Z$_{\odot}$, in the sense that it cannot change the fragmentation behavior for these metallicities. Our results support the idea that low mass fragments can form in the absence of metals, and clouds with $Z  \lesssim$ Z$_{\rm crit}$ will form a cluster with a flat IMF.

\acknowledgments
{We thank Tom Abel, Volker Bromm, Kazuyuki Omukai, Raffaella Schneider, Rowan Smith, and Naoki Yoshida for useful comments. The present work is supported by contract research `Internationale Spitzenforschung II' of the \emph{Baden-W\"urttemberg Stiftung} 
(grant P-LS-SPII/18), the German \emph{Bundesministerium f\"ur Bildung und Forschung} via the ASTRONET project STAR FORMAT (grant 05A09VHA), a Frontier grant of Heidelberg University sponsored by the German Excellence Initiative, the International Max Planck Research School for Astronomy and Cosmic Physics at the University of Heidelberg (IMPRS-HD).
All computations described here were performed at the \emph{Leibniz-Rechenzentrum}, National Supercomputer HLRB-II  (\emph{Bayerische Akademie der Wissenschaften}), on the HPC-GPU Cluster Kolob (University of Heidelberg), and on the Ranger cluster at the Texas Advanced Computing Center, as part of project TG-MCA995024.}

\end{document}